\documentclass[journal]{IEEEtran}
\usepackage[utf8]{inputenc}

\usepackage{tikz} 
\usepackage{tikz-3dplot}

\usetikzlibrary {shapes.symbols}
\usepackage{pgfplots}
\usepgfplotslibrary{fillbetween}
\usepackage[intlimits]{amsmath}
\usepackage{amsthm,amsfonts}
\renewcommand{\vec}[1]{{\boldsymbol#1}}
\newcommand{\mat}[1]{{\mathbf{#1}}}

\newcommand{\C}{\mathbb{C}}
\providecommand*{\mrm}[1]{\mathrm{#1}}

\usepackage{graphicx}

\DeclareMathAccent{\ring}{\mathalpha}{operators}{"17}

\providecommand*{\ohm}{\ensuremath{\mrm{\Omega}}}

\providecommand*{\unit}[1]{\ensuremath{\mrm{\,#1}}}
\providecommand*{\eu}{\ensuremath{\mrm{e}}}

\renewcommand{\Re}{\operatorname{Re}}	
\renewcommand{\Im}{\operatorname{Im}}	
\providecommand*{\diff}{\operatorname{d}\!}

\usepackage{multirow}

\newcommand{\ie}{\textit{i.e.}\/, }
\newcommand{\eg}{\textit{e.g.}\/, }

\providecommand*{\ju}{\ensuremath{\mrm{j}}}

\newcommand{\subto}{\mrm{subject\ to}}

\newcommand{\Jm}{\mat{I}}

\newcommand{\herm}{\mrm{H}}

\newcommand{\Ev}{\vec{E}}
\newcommand{\Jv}{\vec{J}}

\newcommand{\rv}{\vec{r}}

\newcommand{\zvh}{\hat{\vec{z}}}

\newcommand{\reg}{\varOmega}

\definecolor{metal}{RGB}{255, 184, 28}
\definecolor{cut}{RGB}{255,255,255}
\definecolor{diel}{RGB}{45,200,45}

\title{Radiation Efficiency and Gain Bounds \\ for Microstrip Patch Antennas}

\author{\IEEEauthorblockN{
Ben A.P. Nel, Anja K. Skrivervik, and Mats Gustafsson}

\thanks{ 
Corresponding author: Ben A.P. Nel.

This work was supported by Excellence Center at Linköping – Lund in Information Technology (ELLIIT) and the Hedda Andersson guest professor program at Lund University. 

Ben A.P. Nel and Mats Gustafsson are with the Department of Electrical and Information Technology, Lund University, SE-221 00 Lund, Sweden (e-mail:
\{ben.nel,mats.gustafsson\}@eit.lth.se).

Anja K. Skrivervik is with the Microwave Antenna Group, École Polytechnique Fédérale de Lausanne (EPFL), 1015 Lausanne, Switzerland and  the Department of Electrical and Information Technology, Lund University, SE-221 00 Lund, Sweden (e-mail:
anja.skrivervik@epfl.ch).
}}

\begin{document}

\setcounter{equation}{0}
\setcounter{figure}{0}
\setcounter{section}{0}
\setcounter{page}{1}
\newpage
\pagestyle{headings}
\twocolumn

\maketitle
\begin{abstract}
This paper presents radiation efficiency and gain bounds for microstrip patch antennas. The presented bounds are shown to be good predictors of antenna performance. Using the bounds, patch miniaturization techniques based on high permittivity substrates and geometrical shaping are compared. Further, a semi-analytic model is developed to approximate the bounds. Measurements are used to validate the bounds.  Finally, maximum bandwidth of a microstrip patch antenna is linked to its maximum radiation efficiency.      
\end{abstract}


\IEEEpeerreviewmaketitle

\section{Introduction}

\IEEEPARstart{M}{icrostrip} patch antennas have been widely used for several decades~\cite{Pozar1992a, bhar01, James1989}. Today these antennas can be reliably modeled using commercially available computational electromagnetic software, for example, FEKO~\cite{doe:FEKO2} or CST~\cite{doe:CST}. Making use of these simulation tools, antenna designers are able to determine performance parameters such as radiation efficiency, gain, and bandwidth. 

 Radiation efficiency and gain are key performance metrics to consider in the pursuit of reducing antenna losses. For a given antenna design, radiation efficiency and gain can be computed and then optimized using \eg heuristic methods~\cite{Rahmat-Samii+Michielssen1999, Haupt+Werner2007}. Although this is a reliable optimization approach, drawbacks are that it is time consuming and may obtain local optima far from the optimal performance.  

 Early work computing radiation efficiency and gain bounds can be found in~\cite{Harrington1960}. More recent work has focused on using current optimization to obtain radiation efficiency bounds for arbitrary geometries~\cite{Gustafsson+etal2019,Jelinek+Capek2017}. The motivation for this optimization technique is that optimal currents within a given design region provide a performance limit for all designs within that region~\cite{Nel+etal2023a}. Although, to produce the optimal currents, multiple feeds might be required and therefore may not be feasible for single port antennas. 

The goal of this paper is to aid microstrip patch antenna design by providing maximum radiation efficiency and gain bounds. This is achieved by considering all possible patch geometries within a given design region using current optimization~\cite{Gustafsson+Nordebo2013}. Some classical patch antenna designs are shown to perform close to the bounds both when adding Ohmic losses to the patch region as well as when adding dielectric losses to the substrate. Therefore, practical design information is provided regarding the feasibility of obtaining a desired radiation efficiency as well as a benchmark to asses potential design improvements. The work presented here builds on a previous formulation determining lower Q-factor bounds for microstrip patch antennas~\cite{Nel+etal2023a,Tayli+Gustafsson2016}.

In this paper, a single layer microstrip patch antenna is used, with all currents horizontally on the patch design region. To reduce the computational complexity, the ground plane and dielectric substrate are assumed to be infinite. These assumptions are known to be reasonable for moderately sized ground planes and dielectric substrates~\cite{bhar01}.  These microstrip patch antennas can be fed \eg with a probe feed coming from the ground plane. Note that there exist methods that may be used to improve the radiation efficiency of a self resonant microstrip patch antenna, for instance using a shorting pin or by reducing the size of the ground plane and dielectric substrate~\cite{Skrivervik+etal2001, Fujimoto+etal1988, Lee+Tong2012, Shao+Zhang2021}, which are not considered here. However, the bounds presented here serve as a first canonical case for analysing maximum radiation efficiency for antennas that are in wide use.  

Miniaturization is here considered when the patch design region is reduced below its natural resonance in a free-space setting, \eg around half a free-space wavelength for a rectangular design. Two methods of achieving this are evaluated. These are, increasing the substrate permittivity (increasing the design region electrical size) and/or by shaping the patch by removing metal from the patch design region. The latter can reduce the natural half wavelength (in dielectric substrate)  resonance frequency of the metal design region. This study is required as it is well known that reducing antenna size is a challenge, coming at the cost of radiation efficiency~\cite{ Skrivervik+etal2001,Fujimoto+etal1988, Mosallaei+Sarabandi2004a}.

The remainder of the paper is structured as follows; Section~\ref{sec:model} introduces the microstrip patch antenna model and how to evaluate radiation efficiency and gain. Section~\ref{sec:RE} formulates the procedure to compute radiation efficiency and gain bounds using current optimization. Further, patch design region miniaturization is investigated in Section~\ref{sec:mini}. Then using derived semi-analytic expressions, Section~\ref{sec:scale} discusses bounds scaling for Ohmic and dielectric losses. Section~\ref{sec:SEandDL} provides a link between minimum Q-factor and maximum radiation efficiency. A brief discussion on adding vertical currents between the ground plane and patch antenna is presented in Section~\ref{sec:dis}. The paper is concluded in Section~\ref{sec:con}. Finally the appendix provides steps of how half-wavelength patch measurements or simulations can be used to approximate the radiation efficiency bounds for miniaturized designs. 

\section{Microstrip patch antenna model}\label{sec:model}

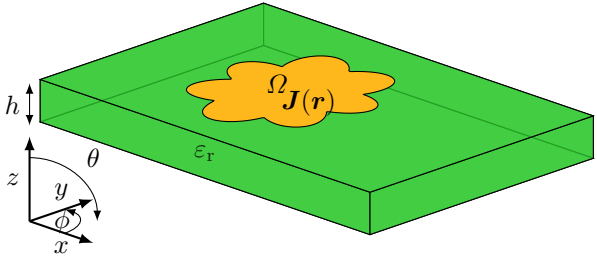
\begin{figure}
\centering
   
   \pgfmathsetmacro{\wx}{2.1} 
\pgfmathsetmacro{\wy}{3.1} 
\pgfmathsetmacro{\wz}{0.3} 
\pgfmathsetmacro{\fx}{0.4} 
\pgfmathsetmacro{\fy}{0}
\pgfmathsetmacro{\fz}{0.6}
\pgfmathsetmacro{\fw}{0.2}
\pgfmathsetmacro{\px}{1} 
\pgfmathsetmacro{\py}{1.5}
\pgfmathsetmacro{\d}{0.2}

\pgfmathsetmacro{\hx}{\px/3} 
\pgfmathsetmacro{\hy}{\py/3}
\pgfmathsetmacro{\Wy}{\py*0.45}
\pgfmathsetmacro{\Lx}{\Wy*0.71}
\pgfmathsetmacro{\Bs}{0.3*\Wy}
\pgfmathsetmacro{\cx}{\px/4} 
\pgfmathsetmacro{\cy}{\py/4}
\pgfmathsetmacro{\lab}{\py/4}
\tikzset{%
		grid/.style={very thin,gray},
		axis/.style={->,white,thin},
		cube/.style={fill=diel,fill opacity=0.8,draw=black},
    patch/.style={fill=metal,fill opacity=1,draw=black},
    patch_cut/.style={fill=cut,fill opacity=1,draw=black},
		cube hidden/.style={fill=diel,fill opacity=0.5,draw=black},
    xyplane/.style={canvas is xy plane at z=#1,very thin},
		>=latex
    }

\tdplotsetmaincoords{70}{135}

\begin{tikzpicture}[tdplot_main_coords]

\begin{scope}[tdplot_main_coords,shift={(9.4,4,2)},scale = 1.2]
    \coordinate (O) at (0,0,0);
    \draw[thick,->] (0,0,0) -- (0,1,0) node[midway,below]{$x$};
    \draw[thick,->] (0,0,0) -- (-1,0,0) node[midway,above]{$y$};
    \draw[thick,->] (0,0,0) -- (0,0,1) node[midway,left]{$z$};

    \draw[->] (0,0.55,0) arc (90:180:0.55) node[midway, left] {$\phi$};

    	\tdplotsetthetaplanecoords{135}
	
	\tdplotdrawarc[->,tdplot_rotated_coords]{(O)}{.75}{0}{90}
          {anchor=south west}{$\theta$}
    \end{scope}

  \draw[cube hidden] (-\wx,-\wy,-\wz) -- (\wx,-\wy,-\wz) -- (\wx,-\wy,\wz) -- (-\wx,-\wy,\wz) -- cycle; 
  \draw[cube hidden] (-\wx,-\wy,-\wz) -- (-\wx,\wy,-\wz) -- (-\wx,\wy,\wz) -- (-\wx,-\wy,\wz) -- cycle; 
	\draw[cube hidden] (-\wx,-\wy,-\wz) -- (-\wx,\wy,-\wz) -- (\wx,\wy,-\wz) -- (\wx,-\wy,-\wz) -- cycle; 
  \begin{scope}[xyplane=-\wz]
  \end{scope}  
	\draw[cube] (\wx,-\wy,-\wz) --
	(\wx,\wy,-\wz) -- (\wx,\wy,\wz) --
		node[below] {$ \varepsilon_\mrm{r}$}
	(\wx,-\wy,\wz) -- cycle; 
	\draw[cube] (-\wx,\wy,-\wz) -- (\wx,\wy,-\wz) -- (\wx,\wy,\wz) -- (-\wx,\wy,\wz) -- cycle; 
	\draw[cube] (-\wx,-\wy,\wz) -- (-\wx,\wy,\wz) -- (\wx,\wy,\wz) -- (\wx,-\wy,\wz) -- cycle;

\begin{scope}[shift={(-0.5,-1,0)},  scale = 0.5]
    \fill[metal] (0,0) -- plot[domain=0:360, samples=180, smooth] ({\x}:{2 + 0.2*cos(2*\x) + abs(0.8*cos(3.5*\x))}) -- cycle;
        \draw[black] plot[domain=0:360, samples=180, smooth] ({\x}:{2 + 0.2*cos(2*\x) + abs(0.8*cos(3.5*\x))});
\end{scope}	   

  \begin{scope}[xyplane=\wz]
	\node at (0,-\py/2) {$\varOmega$};
	\node at (0.2,\py/2-0.7) {$\vec{J}(\vec{r})$};
  \end{scope}  

	\draw[<->] (\wx+\d,-\wy,-\wz) -- node[left] {$h$} (\wx+\d,-\wy,\wz);	
\end{tikzpicture}

   \caption{The microstrip patch antenna design region is given by $\varOmega$. All metal patch geometries fitting within this region are considered.  Surface currents on this design region are denoted, $\vec{J}(\vec{r})$.
   The infinite dielectric substrate with relative permittivity, $\varepsilon_\mrm{r}$, and thickness, $h$, is on top of an infinite PEC ground plane.
   }
   \label{fig:Patch}
\end{figure}

\begin{figure}
\pgfmathsetmacro{\wx}{2.1} 
\pgfmathsetmacro{\wy}{3.1} 
\pgfmathsetmacro{\wz}{0.3} 
\pgfmathsetmacro{\fx}{0.4} 
\pgfmathsetmacro{\fy}{0}
\pgfmathsetmacro{\fz}{0.6}
\pgfmathsetmacro{\fw}{0.2}
\pgfmathsetmacro{\px}{1} 
\pgfmathsetmacro{\py}{1.5}
\pgfmathsetmacro{\d}{0.2}

\pgfmathsetmacro{\hx}{\px/3} 
\pgfmathsetmacro{\hy}{\py/3}
\pgfmathsetmacro{\Wy}{\py*0.45}
\pgfmathsetmacro{\Lx}{\Wy*0.71}
\pgfmathsetmacro{\Bs}{0.3*\Wy}
\pgfmathsetmacro{\cx}{\px/2} 
\pgfmathsetmacro{\cy}{\py/2}
\pgfmathsetmacro{\lab}{\py/4}
\tikzset{%
		grid/.style={very thin,gray},
		axis/.style={->,white,thin},
		cube/.style={fill=diel,fill opacity=0.8,draw=black},
    patch/.style={fill=metal,fill opacity=1,draw=black},
    patch_cut/.style={fill=cut,fill opacity=1,draw=black},
		cube hidden/.style={fill=diel,fill opacity=0.5,draw=black},
    xyplane/.style={canvas is xy plane at z=#1,very thin},
		>=latex
    }

    \tdplotsetmaincoords{0}{90}
\begin{tikzpicture}[tdplot_main_coords]
	\begin{scope}[shift={(0,0,0)},scale = 0.75]
	        	\draw[patch] (-\px,-\py,\wz) -- (-\px,\py,\wz) -- (\px,\py,\wz) -- (\px,-\py,\wz)
	           -- cycle;
	            \node at (0,-\py+\lab,\wz) {(a)};
             \node at (0,-\py+\lab+1,\wz) {$\reg$};
	\end{scope}

  \begin{scope}[shift={(0,3,0)},scale = 0.75]
	        	\draw[patch] (-\px,-\py,\wz) -- (-\px,\py,\wz) -- (\px,\py,\wz) -- (\px,-\py,\wz) -- cycle;

     \draw[patch_cut] (-\cx,0,\wz) -- (-\cx,\cy,\wz) -- (\cx,\cy,\wz) -- (\cx,0,\wz) -- cycle;
				 \node at (0,-\py+\lab,\wz) {(b)};
	\end{scope}

 \begin{scope}[shift={(0,6,0)},scale = 0.75]
\draw[patch] (-\px,-\py,\wz) -- (-\px,-\hy,\wz) -- (-\hx,-\hy,\wz) -- (-\hx,\hy,\wz) -- (-\px,\hy,\wz) -- (-\px,\py,\wz) --(\px,\py,\wz)--
(\px,\hy,\wz)--(\hx,\hy,\wz)--
(\hx,-\hy,\wz)--
(\px,-\hy,\wz)-- (\px,-\py,\wz) -- cycle;
	\node at (0,-\py+\lab,\wz) {(c)};
\end{scope}
 
	\begin{scope}[shift={(0,0,0)},scale = 0.75]
	 \draw[<->] (-\px,-\py-\d,\wz) -- node[left] {$\ell_{\mrm{y}}$} (\px,-\py-\d,\wz);
	 \draw[<->] (\px+\d,-\py,\wz) -- node[below] {$\ell_{\mrm{x}}$} (\px+\d,\py,\wz);
  \end{scope}
  
\end{tikzpicture}
\caption{A rectangular design region $\reg$, see Fig.~\ref{fig:Patch}, having dimensions $\ell_\mrm{x}$ and $\ell_\mrm{y}$ is chosen in this paper. Classical metal patch geometries such as the half-wavelength patch (a), the slot loaded patch (b), and the H-shaped patch (c) fit within this design region.}
\label{fig:geos}
\end{figure}
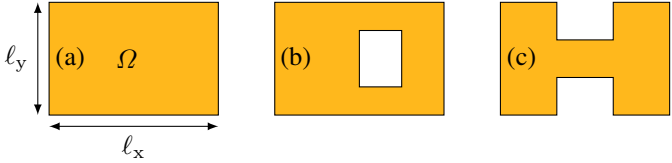

The geometry considered to model microstrip patch antennas is shown in Fig.~\ref{fig:Patch}, where the design region $\varOmega$, that can be of arbitrary shape, is situated at the interface between free space and a transversely infinite dielectric substrate. This dielectric, with relative permittivity $\varepsilon_\mrm{r}$ and thickness $h$, is on top of an infinite PEC ground plane.   

In this paper, a rectangular design region $\varOmega$, is chosen for simplicity, see Fig~\ref{fig:geos}a. Classical patch geometries fitting within this rectangular design region, such as a half-wavelength patch (a) as well as miniaturized geometries that reduce the resonant frequency \eg slot loaded patch~\cite{James1981} (b), and H-shaped patch~\cite{James1981} (c), are shown in Fig.~\ref{fig:geos}. The radiation efficiency and gain of these classical patch geometries serve as a reference with which to compare the presented bounds. It should be noted that these bounds consider all possible patch geometries fitting within the design region $\reg$, thereby obtaining a fundamental limit on achievable maximum radiation efficiency and gain for antenna geometries within $\reg$.

Radiation efficiency and gain are defined as~\cite{Harrington1960} 
\begin{equation}
  \eta =  \frac{P_\mathrm{r}}{P_\mathrm{d}} \quad \text{and } G = 4\pi\frac{U}{P_\mrm{d}},
\label{eq:eff}
\end{equation}
respectively, where $P_\mathrm{r}$ denotes radiated power, $P_\mathrm{d}$ dissipated (accepted) power, and $U$ radiation intensity. For a microstrip patch antenna, the dissipated power can be due to three different loss mechanisms that can be separately analyzed as
\begin{equation}
  P_\mathrm{d} = P_\mathrm{r} + P_\mathrm{\Omega} + P_\mathrm{\varepsilon},
\label{eq:Pd}
\end{equation}
where the Ohmic losses on the patch are given by $P_\ohm$ and losses in the substrate due to dielectric losses and surface waves are given by $P_\mathrm{\varepsilon}$. In reality, the dielectric substrate will always be finite, leading to radiation from surface wave diffraction on the edge. However, this form of radiation is generally undesirable and therefore not considered as such when analyzing radiation efficiency and gain~\cite{Mosig+Gardiol1982}. The remainder of this section focuses on how to analyze the dissipated power, radiation intensity, and radiated power to evaluate the radiation efficiency and gain~\eqref{eq:eff} in a way suitable for current optimization~\cite{Gustafsson+Nordebo2013}. It should be noted that Ohmic losses on the ground plane are not considered in this paper. 

For a given microstrip patch antenna geometry and feed, the total dissipated power ($P_\mathrm{d}$) can obviously be determined from the input voltage and current. However, in this paper all possible geometries on a design region need to be considered and therefore another approach is required. To formulate a current optimization problem, all patch currents need to be related to dissipated power. This can be done using the method of moments (MoM)~\cite{Harrington1968}. By making use of Green's functions, incorporating the effect of the dielectric substrate and the ground plane~\cite{Mosig+Gardiol1982}, the only unknowns of the system are the currents on the design region, see Fig. \ref{fig:Patch}. The current density $\Jv(\rv)$ in the design region $\reg$ is expanded in $N$ basis functions $\vec{\psi}_n(\vec{r})$ as
 \begin{equation}
   \vec{J}(\vec{r}) = \sum_{n = 1}^N I_n \vec{\psi}_n(\vec{r}),
\label{eq:J_r}
\end{equation}
where the position vector is given by $\vec{r}$ with unit vector $\hat{\vec{r}} = \vec{r}/r$ and length  $r = |\vec{r}|$. The MoM impedance matrix $\mat{Z} \in \C^{N \times N}$ relates design region currents to voltages as~\cite{Harrington1968}

\begin{equation}
   \mat{Z}\Jm = \mat{V},
\label{eq:V}
\end{equation}
 where the expansion coefficient $I_n$ are collected in $\Jm \in \C^{N \times 1}$ and excitation voltages in $\mat{V} \in \C^{N \times 1}$.  
 The MoM resistance matrix and reactance matrix are expressed in terms of the impedance matrix as
\begin{equation}
  \mat{R} = \frac{\mat{Z} + \mat{Z}^\herm}{2}  \quad\text{and } \mat{X} = \frac{\mat{Z} - \mat{Z}^\herm}{2 \ju},
\label{eq:Z_mn}
\end{equation}
respectively, where the Hermitian transpose is denoted by superscript $^\herm$ and $\ju^2=-1$. 

The dissipated power~\eqref{eq:Pd} required to obtain radiation efficiency and gain~\eqref{eq:eff} can be computed from the patch currents~\eqref{eq:J_r} and MoM resistance matrix~\eqref{eq:Z_mn} as~\cite{Harrington1968}
\begin{equation}
P_\mathrm{d} = \frac{1}{2}\Jm^\herm\mat{R}\Jm.
\label{eq:R}
\end{equation}
To formulate efficiency and gain optimization problems, it is also required to relate patch currents to radiation intensity and radiated power. This is described in the following subsection.

\subsection{Radiation intensity and radiated power from patch currents}
The current density $\Jv(\rv)$ in the design region is related to the radiated power by integration of the radiation intensity. 
Linked to the radiation intensity is the far field, $\vec{F} = F_\theta \hat{\vec{\theta}} + F_\phi \hat{\vec{\phi}}$, that is defined in terms of the electric field ($r\eu^{\ju k r} \vec{E}$) by letting $r \to \infty$ using spherical
coordinates, where the elevation angle is given by $\theta\in[0,\pi/2)$ and the azimuthal angle is given by $\phi\in[0,2\pi]$,  with the coordinate system shown in Fig.~\ref{fig:Patch}.

The far field contribution from an $\hat{\vec{x}}$-directed  horizontal electric (Hertzian) dipole (HED) with dipole moment $J_\mrm{h}$ (having units $\unit{A m}$) in the design region can be expressed as~\cite{Mosig+Gardiol1982}
\begin{equation}
\begin{aligned}
&  F_\theta =   \frac{Z_0}{2\pi} \frac{- J_\mrm{h}\ju k n_\theta\cos \phi \cos \theta }{n_\theta - \ju \varepsilon_\mrm{r}\cos{\theta} \cot{(k h n_\theta})}\\
& F_\phi = \frac{Z_0}{2\pi}  \frac{J_\mrm{h} \ju k  \sin \phi \cos \theta}{\cos{\theta}- \ju n_\theta\cot{(k hn_\theta})},
\label{eq:asy}
\end{aligned}
\end{equation}
where $Z_0$ is the free space impedance and $n_\theta = \sqrt{\varepsilon_\mrm{r} - \sin^2\theta}$. A simple coordinate rotation can be used to calculate the far field from a $\hat{\vec{y}}$-directed HED. It should be noted that~\eqref{eq:asy} was derived for a lossless dielectric substrate~\cite{Mosig+Gardiol1982} and is here generalized to lossy dielectric substrates. 
To calculate the far field,~\eqref{eq:asy} is extended by integration over the current density ($\Jv$) expanded in basis functions~\eqref{eq:J_r} as
\begin{equation}
 \vec{F}(\mathrm{\theta},\mathrm{\phi}) \approx \mat{F} \mat{I},
\label{eq:ffI}
\end{equation}
where the far-field matrix $\mat{F} \in \C^{2\times N}$ relates patch currents to one far field direction with $\hat{\vec{\theta}}$ and $\hat{\vec{\phi}}$ components.  

The radiation intensity in a direction ($\theta,\phi$) used to evaluate gain in~\eqref{eq:eff} is given by 
\begin{equation}
U =\frac{|\vec{F}|^2}{2Z_0}.
\end{equation}
From this the radiated power can be calculated by integrating
over a half sphere on the free-space side of the design region, where surface wave effects are neglected near the grazing angle ($\theta = \pi/2$) in~\eqref{eq:asy}.
Using a set of quadrature points $(\theta_n,\phi_n)$ together with quadrature weights $w_n$, a matrix $\mat{F}_\mrm{s}$ is constructed by using far-field matrices $\mat{F}$~\eqref{eq:ffI} evaluated at $(\theta_n,\phi_n)$ as rows. For simplicity, square roots for the quadrature weights $w_n$ are incorporated into $\mat{F}_\mrm{s}$ such that the radiated power $P_\mrm{r}$ from patch currents $\Jm$ is determined by a radiation resistance matrix $\mat{R}_\mrm{r} = \mat{F}_\mrm{s}\mat{F}_\mrm{s}$ \ie
\begin{equation}
 P_\mrm{r} =  \frac{1}{2}\Jm^\herm\mat{F}_\mrm{s}^{\herm}\mat{F}_\mrm{s} \Jm = \frac{1}{2} \Jm^\herm\mat{R}_\mrm{r}\Jm.
\label{eq:ff_Rr}
\end{equation}

\section{Bounds on radiation efficiency and gain}~\label{sec:RE}
This section formulates and presents upper bounds on radiation efficiency and gain using current optimization~\cite{Gustafsson+etal2019,Gustafsson+Capek2019}. Maximal efficiency~\eqref{eq:eff} is in the form of a Rayleigh quotient~\cite{Harrington1968}, which can also be written as a quadratically
constrained quadratic program (QCQP)~\cite{Boyd+Vandenberghe2004} as 
\begin{equation}
\begin{aligned}
& \mathrm{max} &&  \Jm^\herm\mat{R}_\mrm{r}\Jm \\
	& \subto && \Jm^\herm\mat{R}\Jm = 2P_\mrm{in}, \\
\end{aligned}  
\label{eq:optnr}
\end{equation}
where the choice of input power $P_\mrm{in}$ does not affect the bounds but only scales the currents. The solution of this optimization problem is in the form of an eigenvalue problem~\cite{Harrington1968} $\eta_\mrm{up} = \max \mrm{eig} (\mat{R}_\mrm{r}, \mat{R})$. 

To enforce self resonance in the radiation efficiency optimization problem~\eqref{eq:optnr}, the reactive power, expressed as a quadratic form over the  reactance matrix ($\mat{X}$)~\eqref{eq:Z_mn}, is set to zero to model a real-valued input impedance ($\Im Z_\mrm{in} = 0$), therefore enforcing self-resonance. This additional constraint reduces the search space of possible optimal currents resulting in the optimization problem
\begin{equation}
\begin{aligned}
& \mathrm{max} &&  \Jm^\herm\mat{R}_\mrm{r}\Jm \\
	& \subto && \Jm^\herm\mat{R}\Jm = 2P_\mrm{in} \\
	&         && \Jm^\herm\mat{X}\Jm = 0.
\end{aligned}  
\label{eq:opt}
\end{equation}
The QCQP~\eqref{eq:opt} can be transformed to a dual problem~\cite{Gustafsson+Capek2019} by multiplying the second constraint with a scalar parameter $\nu$ and adding the two constraints $\Jm^\herm\nu\mat{X}\Jm = 0$ and $\Jm^\herm\mat{R}\Jm =2P_\mrm{in}$ as
\begin{equation}
\begin{aligned}
& \underset{\nu}{\mathrm{min}} \,\underset{\Jm}{\mathrm{max}} &&  \Jm^\herm\mat{R}_\mrm{r}\Jm \\
	& \subto && \Jm^\herm(\mat{R} + \nu\mat{X})\Jm = 2P_\mrm{in}. 
\end{aligned}  
\label{eq:opt_2}
\end{equation}
The dual problem is in the form of a Rayleigh quotient~\cite{Gustafsson+etal2019} and solved as a parametrized eigenvalue problem
\begin{equation}
\eta_\mathrm{ub} =  \underset{\nu}{\mathrm{min}} \,\mathrm{max} \, \mathrm{eig}\big(\mat{R}_\mrm{r},\mat{R} + \nu\mathbf{X}\big),
\label{eq:up}
\end{equation}
where, given the condition $\mat{R} + \nu\mathbf{X} 	\succeq \mat{0}$ and an indefinite $\mat{X}$, the scalar parameter value is restricted to the range 
\begin{equation}
\frac{-1}{\mathrm{max}\,\xi_\mrm{a}} \leq \nu \leq \frac{-1}{\mathrm{min}\,\xi_\mrm{a}},
\label{eq:nu2}
\end{equation}
where $\xi_\mrm{a} = \mathrm{eig}(\mathbf{X},\mat{R})$, resembles characteristic modes~\cite{Harrington+Mautz1971}. Then, the far-field matrix $\mat{F}_\mrm{s}$~\eqref{eq:ff_Rr} is used to rewrite~\eqref{eq:opt_2} as an ordinary eigenvalue problem 
\begin{equation}
\eta_\mathrm{ub} =  \underset{\nu}{\mathrm{min}} \,\mathrm{max} \, \mathrm{eig}\big(\mat{F}_\mrm{s}(\mat{R} + \nu\mathbf{X})^{-1}\mat{F}_\mrm{s}^\herm\big).
\label{eq:up2}
\end{equation}
It should be noted that a simultaneous diagonalization of $\mat{R}$ and $\mat{X}$ can be used to reduce the computational complexity in~\eqref{eq:up2} by inverting a diagonal matrix~\cite{Gustafsson+Capek2019}. 
When recovering the currents from the eigenvalue problem, there may be degenerate eigenvalues that can be handled as shown in~\cite{Capek+etal2017b}. When the currents satisfy the constraints~\eqref{eq:opt}, this means equality in~\eqref{eq:up} and therefore no dual gap~\cite{Beck+Eldar2006} as is generally true for the QCQPs that are presented here with one or two quadratic constraints~\cite{Beck+Eldar2006}.

\begin{figure}
\centering
   \includegraphics[width=\columnwidth]{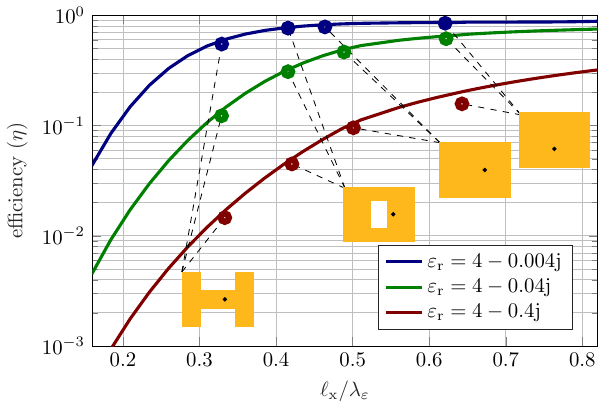}
\caption{Upper bounds on radiation efficiency for all PEC microstrip patch antennas fitting within a rectangular design region $\varOmega$ with dimensions $\ell_\mrm{y} = 0.77 \ell_\mrm{x}$ (see Fig.~\ref{fig:geos}) having specified dielectric loss tangents and height $h = 0.05\ell_\mrm{x}$ (see Fig.~\ref{fig:Patch}). Radiation efficiencies computed using FEKO are shown by markers for the indicated patch antenna geometries (see inset a-c in Fig.~\ref{fig:geos}). }
\label{fig:die_loss}
\end{figure}

 Fig.~\ref{fig:die_loss} shows upper bounds on the radiation efficiency computed for PEC microstrip patch antennas with a dielectric substrate having $\Re\{\varepsilon_\mrm{r}\}=4$ and loss tangent $\tan\delta \in \{0.001,0.01,0.1\}$ using~\eqref{eq:up2}. The design region $\varOmega$ has dimensions $\ell_\mrm{y} = 0.77\ell_\mrm{x}$ and substrate thickness $h= 0.05\ell_\mrm{x}$ (see Fig.~\ref{fig:Patch}), where again it should be noted that the bounds provide a performance limit for all possible patch geometries within the design region. The bounds are shown for a varying patch length $\ell_\mrm{x}$ normalized by the dielectric wavelength, $\lambda_\varepsilon = \lambda /\sqrt{\Re \varepsilon_\mrm{r}}$ (neglecting the imaginary part of permittivity). The bounds are tight from a practical point of view, as shown by the comparison with realistic antennas simulated using commercial software (FEKO) with an infinite ground plane and indicated by the markers to having near optimal performance.  For instance, the half-wavelength patch (see Fig.~\ref{fig:geos}a) is shown to be essentially on the bounds for all presented loss tangents with only a slight deviation from the bounds when exciting the half-wavelength resonance along the shorter dimension ($\ell_\mrm{y}$) and increasing the loss tangent to $\tan{\delta} = 0.1$, however with a different feed this deviation may be reduced. Considering miniaturized geometries, performance near the bounds is also observed for the slot loaded patch (see Fig.~\ref{fig:geos}b) as well as the H-shaped patch (see Fig.~\ref{fig:geos}c). 
 
 The results in Fig.~\ref{fig:die_loss}, as expected show that when the dielectric loss tangent is decreased or the electrical size is increased, the maximum radiation efficiency increases. For design regions smaller than half wavelength in the dielectric ($\ell_\mrm{x}/\lambda_\varepsilon < 0.5$), the bounds show that for high loss tangent \eg $\tan{\delta} = 0.1$ miniaturized designs perform relatively poorly. To demonstrate this compare the bounds for $\tan{\delta} = 0.1$ at $\ell_\mrm{x}/\lambda_\epsilon \approx 1/3$ (realized by H-shaped patch) around 1.5\% efficiency and $\ell_\mrm{x}/\lambda_\epsilon \approx 1/2$ (realized by half wavelength patch) around 10\% efficiency. For a loss tangent $\tan{\delta} = 0.001$ the same comparison leads to bounds at $\ell_\mrm{x}/\lambda_\epsilon \approx 1/3$ of around 60\% and at $\ell_\mrm{x}/\lambda_\epsilon \approx 1/2$ around 80\%. This emphasizes the importance of choosing a substrate with low dielectric losses when considering antennas that are smaller than half a wavelength in the dielectric. It should be noted that further improvements in radiation can be achieved using designs larger than $0.5<\ell_\mrm{x}/\lambda_\varepsilon$. This is done by exciting a half wavelength resonance on the shorter dimension ($\ell_\mrm{y}$) as is well known. Here this is confirmed by the bounds and simulated patch antennas.      

\begin{figure}
\centering
   \includegraphics[width=\columnwidth]{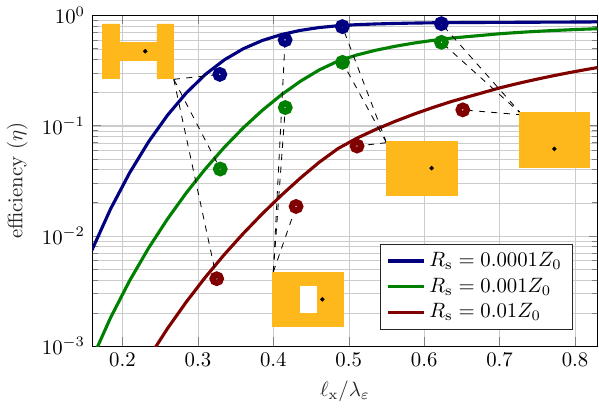}
\caption{Upper bounds on the radiation efficiency for all microstrip patch antennas fitting within the design region with dimensions $\ell_\mrm{y} = 0.77 \ell_\mrm{x}$, specified Ohmic losses, relative permittivity $\varepsilon_\mrm{r} = 4$, and $h = 0.05\ell_\mrm{x}$. Radiation efficiencies for the classical patches shown in Fig.~\ref{fig:geos}  computed using FEKO are shown by markers.}
\label{fig:ohm_loss}
\end{figure}

 Radiation efficiency bounds computed using~\eqref{eq:up2} for patches with varying surface resistivity $R_{\mrm{s}}\in\{0.0377, 0.377, 3.77\}\unit{\Omega/\square}$ (ohms per square) and a lossless dielectric are shown in Fig.~\ref{fig:ohm_loss}.  The surface resistivity can model either a resistive sheet or solid with a skin depth~\cite{Knott+Shaeffer+Tuley2004}. To demonstrate the practical application of the bounds, the performance of different patch designs (see Fig.~\ref{fig:geos}), are compared to the bounds using a commercial software. This shows that the half-wavelength-patches (see Fig.~\ref{fig:geos}a) are essentially on the bounds and therefore optimal for these surface resistivities. Further, the slot loaded patch (see Fig.~\ref{fig:geos}b) and H-shaped patch (see Fig.~\ref{fig:geos}c) perform near the bounds. The reason for the slot loaded patch and the H-shaped patch not being on top of the bounds is attributed to the currents being squeezed for these designs, thus making a less efficient use of the available surface. Similar to Fig.~\ref{fig:die_loss} it is observed that increasing design region dimensions may be used to compensate for high losses.  

A further parameter of interest is gain~\eqref{eq:eff}. Upper bounds on gain can be determined by maximization of the radiation intensity and written as the current optimization problem
\begin{equation}
\begin{aligned}
& \mathrm{max} &&  \Jm^\herm\mat{F}^\herm\mat{F}\Jm \\
	& \subto && \Jm^\herm\mat{R}\Jm = 2 P_\mrm{in} \\
 &  && \Jm^\herm\mat{X}\Jm = 0,
\end{aligned}  
\label{eq:optG}
\end{equation}
which can be formulated as an eigenvalue problem using the range computed for $\nu$ in~\eqref{eq:nu2} as~\cite{Gustafsson+Capek2019} 
\begin{equation}
    G_\mrm{ub,r} \approx 4\pi\ \underset{\nu}{\mathrm{min}} \,\mathrm{max} \, \mathrm{eig}\big(\mat{F}(\mat{R} + \nu\mathbf{X})^{-1}\mat{F}^\herm\big).
    \label{eq:Gopt}
\end{equation}

\begin{figure}
\centering
   \includegraphics[width=\columnwidth]{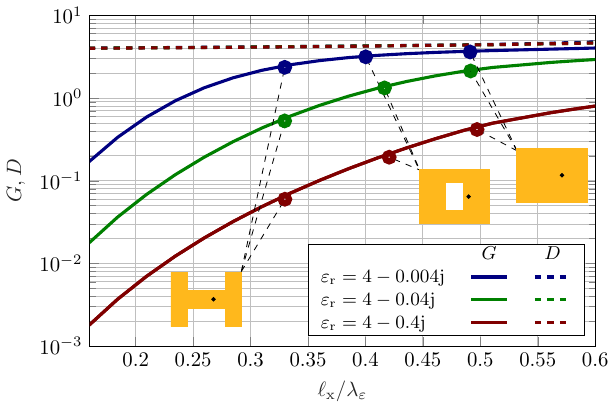}
\caption{Upper bound on the gain in the normal direction, $\zvh$, for microstrip patch antenna, with relative permittivity $\Re \{\varepsilon_\mrm{r}\} = 4$ having varying loss tangent, fitting within a design region with dimensions $\ell_\mrm{y} = 0.77 \ell_\mrm{x}$, and $h = 0.05\ell_\mrm{x}$. The corresponding directivities are shown by dashed lines.  }
\label{fig:Gub}
\end{figure}

Upper bounds on gain for microstrip patch antennas with relative permittivity having real part $\Re\{\varepsilon_\mrm{r}\} = 4$, varying loss tangent, substrate thickness $h = 0.05\ell_\mrm{x}$  and design region dimensions $\ell_\mrm{y} = 0.77\ell_\mrm{x}$ are shown in Fig.~\ref{fig:Gub}. The bounds show that, as expected, an increased dielectric loss tangent reduces the gain. Further, the corresponding directivity ($D = G/\eta$) values, shown by dashed lines, are on top of one another with only a weak dependence on the electrical size of the design region. This suggests that the maximum achievable gain is scaled by achievable radiation efficiency with directivity relatively unaffected. The gain bounds on Ohmic losses are not shown here, however, they lead to the same conclusion on directivity as the bounds with dielectric losses presented in Fig.~\ref{fig:Gub}. This suggests that directivity is mostly determined by the electrical size of the structure when maximizing gain. \footnote{Solving~\eqref{eq:Gopt} may present some challenges recovering the optimal currents from the eigenvectors to test for no dual gap or recover the directivity. To avoid this, the first constraint in~\eqref{eq:opt_2} can be added to the objective and then rewritten as an eigenvalue problem similar to~\eqref{eq:up}. It should be noted that, for all optimization problems presented here, the problem can be reformulated to only search for solutions with radiating currents to improve numerical stability. Further, using semidefinite programming (SDP) the bounds can be computed in an alternative way~\cite{Boyd+Vandenberghe2004}.}

\section{Substrate impact on antenna miniaturization}\label{sec:mini}
In this section, the trade-off between antenna miniaturization and radiation efficiency bounds is investigated. Miniaturizing patch antennas can be achieved by increasing the substrate relative permittivity and/or shaping the patch geometry. A comparison between these two approaches for a given design region and free-space wavelength is provided here. This is done for a design region with dimensions $\ell_\mrm{y} = 0.77\ell_\mrm{x}$, substrate thickness $h = 0.05\ell_\mrm{x}$, and relative permittivities $\Re\{\varepsilon_\mrm{r}\} 
 = 2$ and $\Re\{\varepsilon_\mrm{r}\} 
 = 4$, investigating Ohmic as well as dielectric losses separately. In addition, the radiation efficiency bounds are compared to measurements of miniaturized patch antenna for a given substrate.

\begin{figure}
\centering
   \includegraphics[width=\columnwidth]{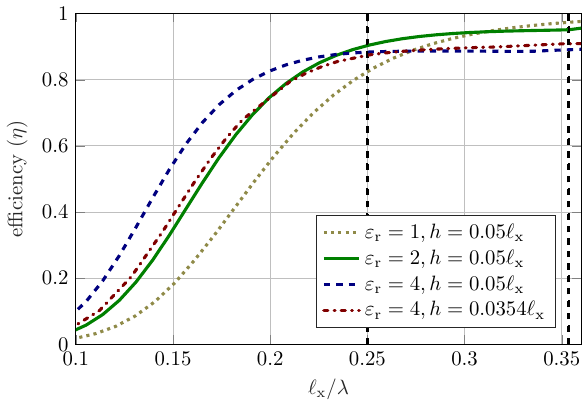}
\caption{Upper bounds on radiation efficiency with surface resistivity $R_\mrm{s} = 0.01 \unit{\Omega/\square}$.  The design region has dimensions $\ell_\mrm{y} = 0.77\ell_\mrm{x}$ and the substrate thickness and relative permittivity indicated in legend. The vertical dashed lines indicate the half wavelength size for the relative permittivity 2 and 4 substrates.} 
\label{fig:Rs24}
\end{figure}

A comparison of the efficiency bounds for different substrate relative permittivities is shown in Fig.~\ref{fig:Rs24} for patch Ohmic losses of $R_\mrm{s} = 0.01 \unit{\Omega/\square}$ (similar to copper) and lossless dielectrics. It is evident that the higher relative permittivity substrate ($\varepsilon_\mrm{r} = 4$) has greater radiation efficiency bounds than the lower permittivity ($\varepsilon_\mrm{r} = 1$ and $\varepsilon_\mrm{r} = 2$) substrates, for structures miniaturized below its half a wavelength (in the $\varepsilon_\mrm{r} = 4$ substrate) $\ell_\mrm{x}/\lambda < 0.25$. The enhanced efficiency of the high permittivity substrate ($\varepsilon_\mrm{r} = 4$) can be attributed in part to its greater electrical thickness. This is demonstrated by comparing its efficiency bound with that of a thinner substrate, $h = 0.0354\ell_\mrm{x}$, representing the electrical thickness of the $\varepsilon_\mrm{r} = 2$ material. The findings suggest that substrates with equal electrical thickness exhibit similar efficiency bounds for $\ell_\mrm{x}/\lambda < 0.25$.

The performance differences among the various substrates become less pronounced for $\ell_\mrm{x}/\lambda > 0.25$, where all scenarios exhibit efficiencies ranging from around $85\%$ to $95\%$. The $\varepsilon_\mrm{r} = 2$ substrate demonstrates marginally better performance up to $\ell_\mrm{x}/\lambda \approx 0.3$, beyond which the free-space scenario ($\varepsilon_\mrm{r} = 1$) excels. In this range, efficiency bounds for the two $\varepsilon_\mrm{r} = 4$ substrates are comparable, with slightly better performance observed for the thinner substrate. The intricate behavior within this range arises from a combination of increased electrical size and increased surface wave losses for the higher permittivity substrates.

\begin{figure}
\centering
   \includegraphics[width=\columnwidth]{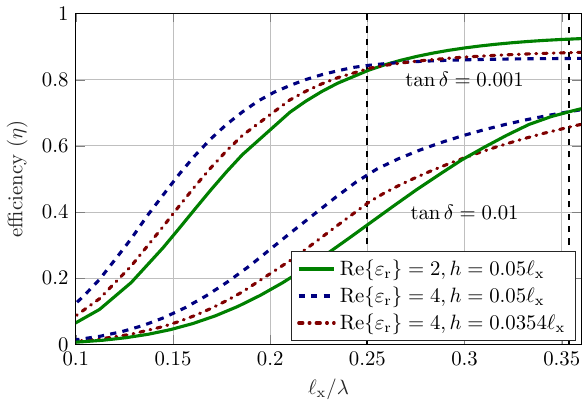}
\caption{Upper bounds on radiation efficiency for two relative permittivities with substrate loss tangents $\tan\delta\in[0.01,0.001]$. The design region has dimensions $\ell_\mrm{y} = 0.77\ell_\mrm{x}$ and indicated substrate thickness. The vertical lines show    }
\label{fig:eps24}
\end{figure}

Further, considering only dielectric losses, the radiation efficiency bounds for relative permittivity $\varepsilon_\mrm{r} = 2(1- \ju \tan\delta)$ and $\varepsilon_\mrm{r} = 4(1- \ju \tan\delta)$ substrates are compared as shown in Fig.~\ref{fig:eps24}. The interpretation remains consistent for the low-loss substrate with $\tan{\delta} = 0.001$, similar to the Ohmic losses depicted in Fig.~\ref{fig:Rs24}. Specifically, the high permittivity substrate ($\Re\{\varepsilon_\mrm{r}\} = 4$) exhibits the highest efficiency bound below $\ell_\mrm{x}/\lambda < 0.25$, while the lower permittivity substrate ($\Re\{\varepsilon_\mrm{r}\} = 2$) outperforms it above $\ell_\mrm{x}/\lambda > 0.25$. Once more, this phenomenon primarily stems from the larger electrical size in electrically small cases and the increased surface wave power for larger sizes. Introducing compensation with an electrical thickness of $h = 0.0354\ell_\mrm{x}$ reduces the efficiency, bringing the values closer to those of the $\Re\{\varepsilon_\mrm{r}\} = 2$ case, although with a disparity larger than observed for the Ohmic losses in Fig.~\ref{fig:Rs24}.

In the case of the more lossy substrate with $\tan{\delta} = 0.01$, the higher permittivity substrate exhibits superior performance across the entire range $\ell_\mrm{x}/\lambda < 0.36$. This can be attributed in part to the dominance of material losses, which hide the impact of surface waves, resulting in lower efficiency overall. Furthermore, the enhancement observed beyond the dielectric's half a wavelength, $\ell_\mrm{x}/\lambda > 0.25$, is attributable to the fact that the $\ell_{\mrm{y}}$-direction becomes half a wavelength increasing the width of the patch.

From these investigations, it can be suggested that when considering miniaturization to choose a higher permittivity substrate over shaping the design region. The exact range over which it remains favorable to do so depends on several factors such as surface wave losses. From the cases considered here it is observed that for low loss cases it is favorable to miniaturize by increasing relative permittivity up to around half a wavelength. To better understand this a further study of how the surface wave depends on design parameters is presented in Section~\ref{sec:scale}.

\begin{figure}
\centering
   \includegraphics[width=\columnwidth]{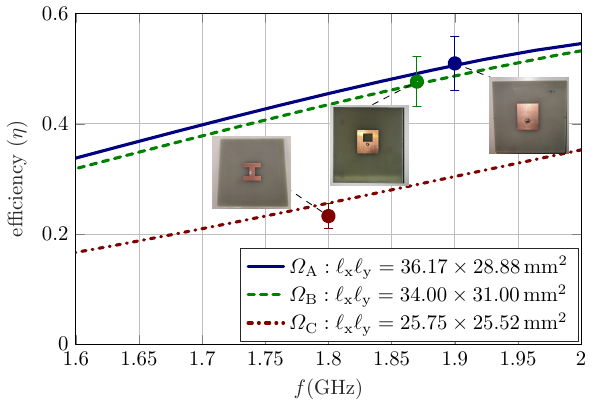}
\caption{Upper bounds on radiation efficiency compared with measurement. The efficiency of three measured antennas are shown with markers. The design regions dimensions are given in the legend. The relative permittivity $h = 3.3\unit{mm}$ thick substrate is $\varepsilon_\mrm{r} = 4.29(1- \ju 0.015)$ and surface resistivity of copper $R_\mrm{s} = 0.01\unit{\Omega/\square}$. Error bars are based on the precision of the gain measurement. } 
\label{fig:meas}
\end{figure}

For a given dielectric substrate, the bounds can be compared to measured antenna designs. This is both a validation of the bounds as well as an investigation of miniaturized designs on the same dielectric substrate. The FR4 dielectric substrate chosen has dimensions $100\unit{mm}\times100\unit{mm}$ with thickness $3.3\unit{mm}$ and relative permittivity $\varepsilon_\mrm{r} \approx 4.29(1- \ju 0.015)$, based on material characterization at $2~\unit{GHz}$ ~\cite{Su2023}. Three different design regions are considered and their radiation efficiency bounds determined between $1.6\unit{GHz}-2\unit{GHz}$ are indicated in Fig.~\ref{fig:meas}. The smallest design region ($\varOmega_\mrm{C}$) has significantly lower radiation efficiency bounds than the other two design regions ($\varOmega_\mrm{A}$ and $\varOmega_\mrm{B}$). When comparing the two largest design regions, the one with the largest maximum dimension has higher radiation efficiency bounds, confirming that for miniaturized geometries it is generally favorable to increase the largest dimension.

Since a finite ground plane is used, part of the undesired surface wave power is detected as radiated power in the measurement. However, since the ground plane is relatively large compared to the design region, the bounds should still be a good indication of optimal performance~\cite{bhar01}. Comparing the half-wavelength patch with design dimensions $\varOmega_\mrm{A}$ at $1.9\unit{GHz}$ with the bounds, the radiation efficiency measured is shown to be very close to the bounds. Similarly, for the slot loaded patch (designed within the design region $\varOmega_\mrm{B}$) the measured radiation efficiency is close to the bounds at $1.87\unit{GHz}$. The H-shaped patch (designed within the design region $\varOmega_\mrm{C}$) has a slight deviation from the bounds at $1.8\unit{GHz}$.

\section{Semi-analytic approximation of bounds}\label{sec:scale}
In this section the contribution of substrate and patch loss parameters pertaining to 
 radiation efficiency bounds are further investigated, to see if this effect can be approximated to enhance understanding and simplify computations. With this, insights can be derived as to how changes to the substrate or patch material properties affect the radiation efficiency bounds.

 The dissipation factor~\cite{Harrington1960,Gustafsson+etal2019}, defined as
\begin{equation}
    \varDelta = \frac{P_\mrm{d} - P_\mrm{r}}{P_\mrm{r}}
    =\frac{P_\varepsilon+P_\Omega}{P_\mrm{r}}
    \label{eq:df}
\end{equation}
is a natural parameter to consider when investigating the effect of loss parameters on scaling of radiation efficiency bounds (obtained with~\eqref{eq:up2}).
The dissipation factor is related to the radiation efficiency as $\eta = (1 + \varDelta)^{-1}$. Therefore, upper bounds on radiation efficiency provide lower bounds on dissipation factor. 

\begin{figure}
\centering
  \includegraphics[width=\columnwidth]{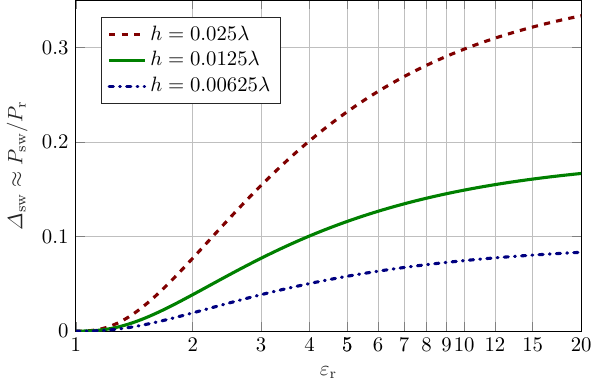}
\caption{Ratio between the first surface wave mode power ($P_\mrm{sw}$) and radiated power ($P_\mrm{r}$) for a HED approximated by~\eqref{eq:rad_guide}. The expression is evaluated over a rage of relative permittivities and for three electrical thicknesses. } 
\label{fig:swratio}
\end{figure}

As shown in~\cite{Gustafsson+etal2019} for antennas
in free space, lower bounds on dissipation factor scales linearly with surface resistivity ($R_\mrm{s}$).  However, only considering this scaling for microstrip patch antennas does not account for surface wave effects, as, even with lossless materials, $P_\varepsilon \neq 0$. To account for this effect, an approximate expression~\cite{Mosig1989}    
\begin{equation}
\frac{P_\mathrm{sw}}{P_\mathrm{r}} \approx \varDelta_\mrm{sw} 
=    \frac{3\pi}{4} \frac{\left(\Re\{\varepsilon_\mrm{r}\}-1\right)^3k h}{\Re\{\varepsilon_\mrm{r}\}^2\left(\Re\{\varepsilon_\mrm{r}\} -1\right)+\frac{2}{5}\Re\{\varepsilon_\mrm{r}\}}
\label{eq:rad_guide}
\end{equation} 
relating the surface wave power to the radiated power of an HED is used, see Fig.~\ref{fig:swratio}.
The surface wave is strictly only defined for lossless dielectrics but in~\eqref{eq:rad_guide} it is assumed that the ratio between propagated power in the dielectric and radiated power~\eqref{eq:ff_Rr} remains constant with increased loss tangent. Further, due to the choice of thin dielectric substrates, only the first transverse magnetic surface wave mode is propagating~\cite{Mosig1978}. The thickness required to have the first transverse electric surface wave mode propagating in the substrate is $h > \lambda/(4 \sqrt{\varepsilon_\mrm{r} - 1})$~\cite{Mosig+Gardiol1982} in a lossless substrate. 

For three electrical thicknesses over a range of relative permittivities the expression~\eqref{eq:rad_guide} is used to approximate the surface wave to radiated power ratio as shown in Fig.~\ref{fig:swratio}. In the figure the $h = 0.0125\lambda$ at $\varepsilon_\mrm{r} = 4$ corresponds to a relative permittivity of 4 substrate at half a dielectric wavelength in Fig.~\ref{fig:die_loss} and Fig.~\ref{fig:ohm_loss}. It should be noted that by increasing the relative permittivity higher order surface wave modes could be excited although this is not the case in Fig.~\ref{fig:swratio}.

\begin{figure}
\centering
   \includegraphics[width=\columnwidth]{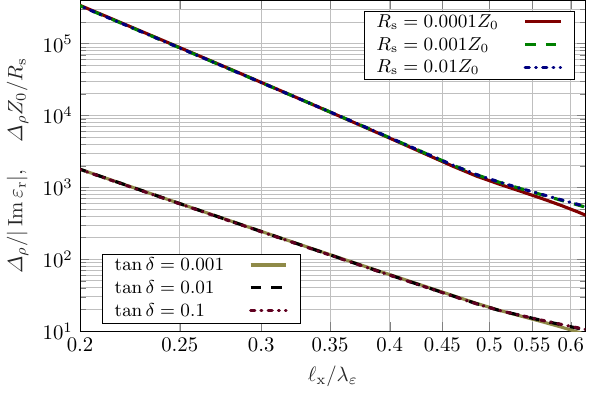}
\caption{Lower dissipation factor bounds normalized by surface resistivity summed with the relative  permittivity with the approximate surface wave effect~\eqref{eq:rad_guide} subtracted shown for a microstrip patch
antenna with relative permittivity  $\Re\{\varepsilon_\mrm{r}\} = 4$, dimensions $h = 0.05\ell_\mrm{x}$ and $\ell_\mrm{y} = 0.77 \ell_\mrm{x}$.}
\label{fig:norm_ohm_sw}
\end{figure}

The normalized surface wave power~\eqref{eq:rad_guide} can be approximately removed from the dissipation factor~\eqref{eq:df} as
\begin{equation}
    \varDelta_\rho = \varDelta - \varDelta_\mrm{sw}.
    \label{eq:no_SW}
\end{equation}
This is applied to the lower dissipation factor bounds, along with normalizing with the~\eqref{eq:up2} surface resistivity ($\varDelta_\rho Z_0/R_\mrm{s}$), as shown in Fig.~\ref{fig:norm_ohm_sw}. The results show very little difference between the bounds for the different resistivities $R_{\mrm{s}}$ when removing the two main contributions (surface resistivity and surface wave). It is noted that as the electrical size increases, the bounds start to deviate slightly. This is due to the surface wave approximation~\eqref{eq:rad_guide} not being accurate for these cases~\cite{Mosig1989} along with the optimal currents trying to suppress these losses as they become more significant. It is very important to note that~\eqref{eq:rad_guide} does not indicate a bound on the ratio between surface wave and radiated power. However, it is an excellent approximation when the surface wave power is not a dominant contribution to radiation efficiency as confirmed in Fig.~\ref{fig:norm_ohm_sw}.

To investigate the bounds' dependence on dielectric losses, the relationship between dissipated power in the dielectric substrate and the imaginary part of the relative permittivity is required. This is given by
\begin{equation}
P_\mrm{\varepsilon} = \frac{-1}{2} \int \Im \{\varepsilon_\mrm{r}(\rv)\}|\Ev(\rv)|^2 \diff\mathrm{V},
\label{eq:P_die}
\end{equation}
where it is sufficient to integrate over the volume of the substrate. Assuming that the non propagating (excluding surface waves) part of the electric field in the substrate remains constant when the dielectric losses are increased, then the dissipation factor~\eqref{eq:df} should scale linearly with respect to the dielectric losses~\eqref{eq:P_die}, neglecting the surface wave. Further, as only the real part of the relative permittivity is taken in~\eqref{eq:rad_guide}, the ratio between radiated and surface wave power is assumed to remain constant. 

The normalized lower bounds on dissipation factor for varying dielectric loss tangent are shown in Fig.~\ref{fig:norm_ohm_sw}. The bounds with this normalization, along with subtracting the surface wave power~\eqref{eq:rad_guide}, are approximately equal for all presented loss tangents. This means that the lower dissipation factor bounds scale approximately linearly with respect to loss tangent after the approximate surface wave  power is removed, implying negligible changes to the near field. Further, the ratio of surface wave power to radiated power approximately follows~\eqref{eq:rad_guide} except for electrically large patches.

\section{Closed form expression of Maximum radiation efficiency linked to minimum Q-factor}\label{sec:SEandDL}
In this section, a link between stored electric energy in the substrate and dielectric losses is established. This allows for microstrip patch antenna Q-factor to be related to its radiation efficiency leading to a link between maximum radiation efficiency and minimum Q-factor (maximum bandwidth).  This link requires that most of the stored electric energy is confined in the dielectric substrate similar to the assumption made in the cavity model~\cite{James1989}. With this assumption, the stored electric energy can be related to the dissipated power in the near field (due to dielectric substrate losses) and along with surface wave power~\eqref{eq:rad_guide} can be related to the total dissipated power in the substrate as
\begin{equation}
    P_\varepsilon \approx 2\omega W_\mrm{e} \tan \delta + P_\mrm{sw}, 
    \label{eq:P_eps}
\end{equation}
where the stored electric energy is given by $W_\mrm{e}$.  
Adding the radiated power to~\eqref{eq:P_eps} and using the total dissipated power~\eqref{eq:Pd}, leads to 
\begin{equation}
    P_\mrm{d} \approx  P_\mrm{r} + P_\mrm{sw} + 2\omega W_\mrm{e} \tan \delta,
    \label{eq:eta_we}
\end{equation}
assuming no Ohmic losses. It is useful to rewrite~\eqref{eq:eta_we} in terms of Q-factor that can be approximated from fractional bandwidth or input impedance frequency derivative~\cite{Yaghjian+Best2005}. This can easily be measured with \eg a VNA. 

Self resonant antennas have equal stored electric and magnetic energies, which simplifies the Q-factor~\cite{Yaghjian+Best2005} to 
\begin{equation}
Q = \frac{2\omega W_\mathrm{e}}{P_\mathrm{d}}.
\label{eq:Q-factor}
\end{equation}
Substituting the Q-factor~\eqref{eq:Q-factor} into~\eqref{eq:eta_we} normalized by dissipated power and identification of the efficiency~\eqref{eq:eff} yields 
\begin{equation}
    1 \approx \eta + \frac{P_\mrm{sw}}{P_\mrm{d}} + Q \tan \delta.
    \label{eq:eta_Q}
\end{equation}
 
 Using the approximation~\eqref{eq:rad_guide} for the surface wave power, the radiation efficiency can be factored out in~\eqref{eq:eta_Q} and expressed as     
\begin{equation}
    \eta \approx \frac{1 - Q \tan \delta}{1 +  \varDelta_\mrm{sw}}.
\label{eq:eta_Q_sw}
\end{equation}
Assuming the ratio of surface wave power to radiated power ($\varDelta_\mrm{sw}$) remains constant when loss tangent is increased, it is clear from~\eqref{eq:eta_Q_sw} that minimizing Q-factor is equivalent to maximizing radiation efficiency. This means that maximizing bandwidth and radiation efficiency is very closely related for self resonant microstrip patch antennas. 

Lower Q-factor bounds of a lossless substrate can be related to maximum radiation efficiency~\cite{walter1975} of a lossy substrate when the radiated Q-factor ($Q/\eta$) is assumed to be invariant with respect to loss tangent as suggested by losses presented in Fig.~\ref{fig:norm_ohm_sw}. The lower Q-factor bounds for a patch antenna on a lossless substrate ($Q_\mrm{lb}$) is here defined in terms of Q-factor of a half wavelength patch ($Q_\mrm{hw}$) with dielectric losses as 
\begin{equation}
    Q_{\mrm{lb}} = \frac{Q_\mrm{hw}}{\eta(1 + \varDelta_\mrm{sw})}
\end{equation}
Substituting this into the expression relating efficiency to Q-factor~\eqref{eq:eta_Q_sw} leads to an approximation of maximum radiation efficiency in terms of the lower Q-factor bound for a patch antenna on a lossless substrate as 
\begin{equation}
\eta_\mrm{ub} \approx  \frac{1}{(Q_\mathrm{lb} \tan{\delta} + 1)(\varDelta_\mrm{sw} + 1)}.
\label{eq:opt_Qlb}
\end{equation}
 The lower Q-factor bounds in this expression can be determined from results presented in~\cite{Nel+etal2023a} by comparing the result with maximum radiation efficiency obtained from~\eqref{eq:up2}.
 
\begin{figure}
\centering
  \includegraphics[width=\columnwidth]{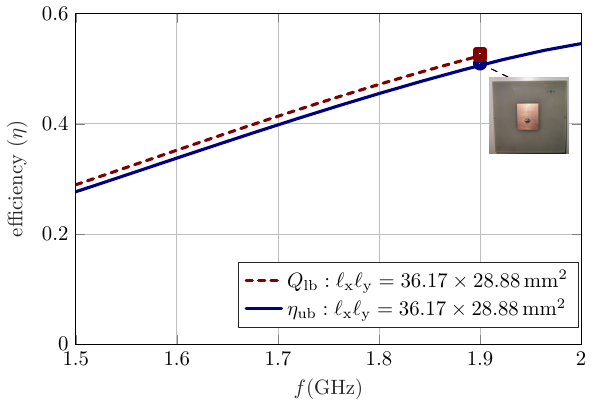}
\caption{Using~\eqref{eq:opt_Qlb} to compute approximate radiation efficiency bounds from a measured microstrip patch antenna with relative permittivity $\varepsilon_\mrm{r} = 4.29(1 - \ju 0.015 )$. The approximate bounds from an approximation of lower Q-factor is compared to the bounds. } 
\label{fig:Pr}
\end{figure}

 The steps required to approximate radiation efficiency bounds from Q-factor of a single half wavelength patch antenna assuming negligible Ohmic losses are outlined in Appendix~\ref{app:Q2re}. This can be useful as Q-factor can be determined from measurements and simulations. The results, as shown in Fig.~\ref{fig:Pr}, indicate that lower Q-factor bounds of a microstrip patch antenna is a good approximation of maximum achievable radiation efficiency when dielectric losses are added. Further, applying a scaling rule the measured results can be used to approximate maximum achievable radiation efficiency for miniaturized designs as shown by the dashed line in Fig.~\ref{fig:Pr}.

In~\cite{James1989} Q-factor is linked to Ohmic losses of half wavelength microstrip patch antennas by using the cavity model approximation. This can be used to write a similar expression to~\eqref{eq:opt_Qlb} for only Ohmic losses by replacing $\tan \delta$ in~\eqref{eq:opt_Qlb} with $2R_\mrm{s}/(khZ_0)$. This assumes Ohmic losses on both the ground plane and patch. The assumption made here of only Ohmic losses on the patch region can be made by replacing with $R_\mrm{s}/(khZ_0)$ instead. This expression is expected to be less accurate than~\eqref{eq:opt_Qlb} as the stored energy is less clearly linked to Ohmic losses in a general setting.

\section{Vertical currents}\label{sec:dis}
\begin{figure}
\centering
   \includegraphics[width=\columnwidth]{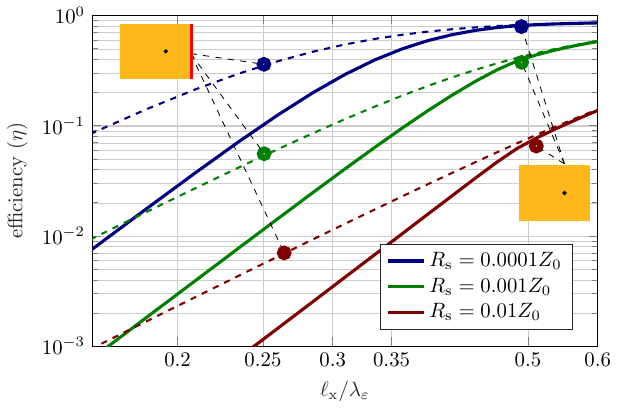}
\caption{Upper bounds on radiation efficiency compared with (solid) and without (dashed) self resonant constraint. The bounds are given for varying surface resistivities. Along with the performance of half wavelength patch antennas some PIFAs radiation efficiency is also shown. The side of the PIFAs shorted to ground is indicated with red.}
\label{fig:pifa_ohm}
\end{figure}

In this paper, the bounds do not account for vertical currents between the ground plane and the dielectric substrate. Based on the bounds presented thus far, using a probe feed produces performance close to the theoretical limits. However, it is worth noting an interesting approach: using a shorting pin/wall to miniaturize the patch. This avoids having to reshape the rectangular metal design region into for instance an H-shaped patch (see Fig.~\ref{fig:geos}c) but requires the addition of vias, leading to planar inverted-F antenna (PIFA) designs. These antennas can be simulated in commercial software using a shorting wall, as shown in Fig.~\ref{fig:pifa_ohm}, demonstrating significantly higher radiation efficiency compared to the self resonant bounds. To better understand
the reason for this discrepancy the self resonant constraint
is removed~\eqref{eq:optnr}, demonstrating significantly higher radiation efficiency compared to the self resonant bounds below half a wavelength. This leading to bounds performance similar to
the PIFA radiation efficiency. Suggesting that PIFA antennas essentially make the optimal Ohmic loss patch currents
self resonant. It should be noted that the PIFA also performs better than the self resonant dielectric loss bounds. However,
when the self resonant constraint is removed, highly inductive
loop currents that do not radiate in the normal direction and produce high Ohmic losses, lead to significantly higher radiation efficiency
bounds.        

\section{Conclusion}\label{sec:con}
 In this paper, radiation efficiency and gain bounds are shown to be excellent predictors of microstrip patch antenna performance. This is demonstrated by comparing the bounds with simulated and measured antennas, considering both Ohmic and/or dielectric losses. That it is possible to construct antennas near the presented bounds is encouraging for antenna designers. However, a conclusion that should not be drawn is that all designs will be near the bounds, to the contrary it would be easy to show that many designs do not have this property.       

One application of the bounds is to assess patch miniaturization designs potential performance. For this, increasing substrate relative permittivity and/or shaping the metal design region are compared. It is found that increasing permittivity is the preferred option for miniaturization up to half a wavelength in substrate with low losses. For higher losses, miniaturization by increasing permittivity becomes even more favorable assuming a constant loss tangent. 

To understand how changes in Ohmic as well as dielectric losses affect the bounds, a semi-analytic approach to the bounds is proposed. This shows that, along with using an approximate expression predicting the surface wave losses, the material parameter effect on the bounds can be approximated. From a design perspective this quantifies the cost of higher material losses in a simple to calculate way.

Along with radiation efficiency, Q-factor, which is inversely proportional to fractional bandwidth, is an important design parameter. Therefore, the link between minimum Q-factor and maximum radiation efficiency is investigated. This leads to the finding that lower Q-factor bounds can be used to approximate maximum radiation efficiency for microstrip patch antennas. The practical benefit of this is twofold. First, the Q-factor (generally easy to measure) can be used to approximate radiation efficiency, and secondly, an antenna optimized for one of these parameters is likely to be near optimal for the other. It should, however, be emphasised that this link is an approximation and ideally both bounds should be computed separately. A Pareto front can be used in the future to study this~\cite{Gustafsson+etal2019, Boyd+Vandenberghe2004}. 

The bounds presented in this paper serve as a first canonical case for benchmarking widely used antennas. The effect of including vertical currents and a finite grounds plane can be investigated in future research. The former, requiring the addition of vias, is shown to be an effective miniaturization technique when compared to the bounds presented in this paper.

\appendices
\section{Relation between Q-factor and radiation efficiency}\label{app:Q2re}
This appendix outlines the steps to go from the Q-factor of a lossy substrate to an approximation of maximum radiation efficiency of a miniaturized design region and provides a practical example. Here the Q-factor is obtained from a half wavelength patch measurement, however, a simulation could also have been used. Since the dielectric losses are relatively high using an FR4 substrate, the Ohmic losses are ignored when going from Q-factor to radiation efficiency in the following example.

To illustrate the practical use of the expressions in section~\ref{sec:SEandDL} on how to use the Q-factor of a half wavelength patch antenna to approximate the maximum achievable radiation efficiency for a miniaturized design region, the following steps are used:   

\begin{enumerate}
    \item Measure Q-factor of half wavelength patch \eg from bandwidth or impedance~\cite{Yaghjian+Best2005}
    \item Approximate the ratio of surface wave to radiated power using~\eqref{eq:rad_guide}
    \item Determine the approximate radiation efficiency using~\eqref{eq:eta_Q_sw} 
    \item Determine the approximate lossless substrate Q-factor using $Q_\mrm{lb} = Q/(1-Q\tan{\delta})$ 
    \item Scale the Q-factor~\cite{Nel+etal2023a}

    \item Compute new approximate surface wave to radiated power ratio using~\eqref{eq:rad_guide}
    \item Convert scaled Q-factor to approximate radiation efficiency bounds using~\eqref{eq:opt_Qlb}
\end{enumerate}

As a practical example, 
using the half wavelength patch in Fig.~\ref{fig:Pr},   resonant at 1.9\unit{GHz} these steps can be demonstrated. This design region has dimensions $\ell_\mrm{x} = 36.17\unit{mm}$ ($\ell_\mrm{x}/\lambda_\varepsilon = 0.476$), $\ell_\mrm{y} = 28.88\unit{mm}$ and $h = 3.3\unit{mm}$. The substrate relative permittivity is $\varepsilon_\mrm{r} = 4.29(1 - \ju 0.015)$. For step 1, the Q-factor of this patch antenna is determined from its fractional bandwidth to be $Q_\mrm{hw} = 25.4$ (approximately the same value can be obtained from the input impedance frequency derivative). Then 2 the surface wave power to radiated power ratio is determined from~\eqref{eq:rad_guide} to be $\varDelta_\mrm{sw} = 0.178$. Using the first two steps, in step 3 the approximate radiation efficiency of the half wavelength patch is calculated from~\eqref{eq:eta_Q_sw}. The resulting approximate efficiency is $\eta \approx 0.526$. This radiation efficiency approximation is added to the bounds shown in Fig.~\ref{fig:Pr} and show a near optimal value compared to the bounds. The next step 4 is to approximately convert the Q-factor with dielectric losses removed. This leads to $Q_\mrm{lb} = 41$. It may be of interest to assess what the expected approximate maximum radiation efficiency will be miniaturizing to $1.5\unit{GHz}$ with the same design parameters. Then in 5, scaling the Q-factor to this frequency, the approximate lower bounds are $Q_\mrm{lb} \approx 135.4$. Now, the approximate surface wave to radiated power is calculated in 6 as $\varDelta_\mrm{sw} = 0.14$. Finally in 7, the approximate achievable radiation efficiency at $1.5\unit{GHz}$ is $\eta \approx 0.29$. The approximated bounds between these two points are approximated by a dashed line. 
\section*{Acknowledgment} \addcontentsline{toc}{section}{Acknowledgment}
We thank Hannes Bartle and Dr. Ismael Triviño from EPFL for helping with measurements.

\end{document}